\documentclass[iop,apj,tighten,revtex4]{emulateapj}
\usepackage{multirow}
\usepackage{amsmath,amssymb,graphics,graphicx}

\begin{document}

\title{Nuclear Thermometers for Classical Novae}

\author{Lori N. Downen}
\author{Christian Iliadis}
\affiliation{Department of Physics and Astronomy, University of North Carolina at Chapel Hill, Chapel Hill, NC 27599-3255, USA}
\affiliation{Triangle Universities Nuclear Laboratory, Durham, NC 27708-0308, USA}
\author{Jordi Jos\'{e}}
\affiliation{Departament de F\'{i}sica i Enginyeria Nuclear, EUETIB, Universitat Polit\`ecnica de Catalunya, E-08036 Barcelona, Spain}
\affiliation{Institut d'Estudis Espacials de Catalunya, E-08034 Barcelona, Spain}
\author{Sumner Starrfield}
\affiliation{School of Earth and Space Exploration, Arizona State University, Tempe, AZ 85287-1404}

%\altaffiltext{3}{Departament de F\'{i}sica i Enginyeria Nuclear, EUETIB, Universitat Polit\`ecnica de Catalunya, E-08036 Barcelona, Spain
%\author{Lori Downen\altaffilmark{1,2}, Christian Iliadis\altaffilmark{1,2}, Jordi Jos\'{e}\altaffilmark{3,4} and Sumner Starrfield\altaffilmark{5}}
%\altaffiltext{1}{Department of Physics and Astronomy, University of North Carolina at Chapel Hill, Chapel Hill, NC 27599-3255, USA}
%\altaffiltext{2}{Triangle Universities Nuclear Laboratory, Durham, NC 27708-0308, USA.}
%\altaffiltext{3}{Departament de F\'{i}sica i Enginyeria Nuclear, EUETIB, Universitat Polit\`ecnica de Catalunya, E-08036 Barcelona, Spain}
%\altaffiltext{4}{Institut d'Estudis Espacials de Catalunya, E-08034 Barcelona, Spain}
%\altaffiltext{5}{School of Earth and Space Exploration, Arizona State University, Tempe, AZ 85287-1404}

\begin{abstract}\label{abstract}
Classical novae are stellar explosions occurring in binary systems, consisting of a white dwarf and a main sequence companion. Thermonuclear runaways on the surface of massive white dwarfs, consisting of oxygen and neon, are believed to reach peak temperatures of several hundred million kelvin. These temperatures are strongly correlated with the underlying white dwarf mass. The observational counterparts of such models are likely associated with outbursts that show strong spectral lines of neon in their shells (neon novae). The goals of this work are to investigate how useful elemental abundances are for constraining the peak temperatures achieved during these outbursts and determine how robust ``nova thermometers" are with respect to uncertain nuclear physics input. We present updated observed abundances in neon novae and perform a series of hydrodynamic simulations for several white dwarf masses. We find that the most useful thermometers, N/O, N/Al, O/S, S/Al, O/Na, Na/Al, O/P, and P/Al, are those with the steepest monotonic dependence on peak temperature. The sensitivity of these thermometers to thermonuclear reaction rate variations is explored using post-processing nucleosynthesis simulations. The ratios N/O, N/Al, O/Na, and Na/Al are robust, meaning they are minimally affected by uncertain rates. However, their dependence on peak temperature is relatively weak. The ratios O/S, S/Al, O/P, and P/Al reveal strong dependences on temperature and the poorly known $^{30}$P(p,$\gamma$)$^{31}$S rate. We compare our model predictions to neon nova observations and obtain the following estimates for the underlying white dwarf masses: 1.34-1.35 M$_{\odot}$ (V838 Her), 1.18-1.21 M$_{\odot}$ (V382 Vel), $\leq$1.3 M$_{\odot}$ (V693 CrA), $\leq$1.2 M$_{\odot}$ (LMC 1990\#1), and $\leq$1.2 M$_{\odot}$ (QU Vul).
\end{abstract}


\section{Introduction\label{intro}}
Classical novae are stellar explosions occurring in binary systems consisting of a white dwarf and a close companion star \citep{MB08}. When the stellar companion expands beyond its Roche Lobe, a mass transfer episode ensues. This matter possesses angular momentum due to the system's rotation and forms an accretion disk, falling onto the surface of the white dwarf. At some point, matter from the underlying white dwarf is mixed with the accreted material, and this mixture is heated and compressed until a thermonuclear runaway takes place, ejecting matter into the interstellar medium. The ejected material consists of a mixture of white dwarf and accreted materials that has been processed by explosive hydrogen burning \citep[and references therein]{SS08,MS10,OY05}. Spectroscopic studies have identified two distinct white dwarf compositions, carbon-oxygen (CO) and oxygen-neon (ONe). Nova shells rich in CNO material point to an underlying CO white dwarf, which represents the evolutionary fate of a low-mass star after core helium burning. On the other hand, element enrichments\footnote{Neon novae are mainly identified by the early appearance of the [Ne II] 12.8 $\micron$ emission line \citep{RG98}.} in the range of Ne to Ar have been attributed to underlying, massive ONe white dwarfs (M$_{WD}\gtrapprox$ 1.1 M$_{\odot}$), representing the evolutionary fate of stars (M $\approx$ 9-11 M$_{\odot}$) after completion of core carbon burning \citep[and references therein]{LA10}. The latter explosions, which are often referred to as {\it neon novae}, tend to be much more energetic than CO novae \citep{SS86} and are the main interest of the present work.
	
The study of classical novae is of significant interest for several reasons. Spectroscopic studies, after proper interpretation, constrain stellar evolutionary models by revealing the composition of the underlying white dwarf. Observed elemental abundances also provide a record of the thermonuclear runaway, including peak temperatures and expansion timescales, and have been used to constrain stellar explosion models \citep{SS00}. Classical novae are noteworthy contributors to the chemical evolution of the Galaxy because they are major sources of $^{13}$C, $^{15}$N, and $^{17}$O  \citep{AK97,JJ98,JJ06,SS08}. Significant efforts have been undertaken to detect $\gamma$-rays from novae as well. Although no nuclear $\gamma$-rays from classical novae have been observed by any satellite observatory \citep{MH08}, it has been proposed that $\gamma$-rays should be present in these novae due to the decay of radioactive $^{18}$F.
	
Although theoretical models of classical novae are successful in reproducing the overall characteristics of the observed outbursts, several key issues remain unexplained and call for improved observations, simulations, and nuclear laboratory measurements. First, the masses of ONe white dwarfs are poorly constrained by observations. Models of intermediate-mass stars predict values in the range of $\approx$ 1.1-1.4 M$_{\odot}$ \citep[see discussion in][]{CD10}. Clearly, accurate white dwarf mass determinations are highly desirable. Second, hydrodynamic simulations underpredict the observed ejecta masses by an order of magnitude or more \citep{SS08}. Third, although it is sometimes claimed in the literature that theoretical models more or less reproduce the observed elemental abundance patterns of classical novae, we will show that such claims are premature regarding neon novae.
	
The goal of the present work is to investigate if and how simulated elemental abundances in classical nova shells can strongly constrain important explosion parameters, most notably the peak temperature. In particular, we seek a robust relationship between model peak temperature and simulated elemental abundance ratios. Such a classical nova {\it thermometer}, if it exists, should be useful for improving models of thermonuclear runaways. Reliable determination of the peak temperature during the outburst will also greatly improve estimates of the mass of the underlying white dwarf.

In this work, we focus on exploring the peak temperature (or, equivalently, white dwarf mass) parameter space of classical nova models. Clearly, the degree of mixing between white dwarf matter and the accumulated matter before the explosion, or more generally, the mixing mechanism \citep[see discussion in][]{JC11}, significantly impacts the ejected abundances. However, in this work we will assume a mixing fraction of 50\% and will leave a study of varying this parameter to future work. 

From a nuclear physics point of view, the differences of the nucleosynthesis pattern in the $A<20$ and $A\geq20$ mass regions are readily apparent. In the former case, the hot CNO cycles give rise to a redistribution of material in C, N, and O and, in favorable cases, the peak temperature is likely reflected in the observed abundance ratios. In the latter case, the nucleosynthesis is characterized by a continuous flow from Ne up, perhaps influenced by reaction cycles at $^{23}$Na, $^{27}$Al, $^{31}$P, and $^{35}$Cl \citep{CI07}.

Thermonuclear reaction rates are highly sensitive to the stellar temperature and, consequently, small temperature changes have a large impact on elemental abundance ratios. This argument represents our starting point for studying potential thermometers for classical novae. For example, $^{20}$Ne(p,$\gamma$)$^{21}$Na is among the slowest nuclear reactions during the thermonuclear runaway. The main reason is the low reaction Q-value, giving rise to a very small number of resonances at low center-of-mass energies. In fact, the large Ne abundance observed in the ejecta of {\it neon novae} is mainly caused by the low proton capture rate on $^{20}$Ne\footnote{The operation of the $^{23}$Na(p,$\alpha$)$^{20}$Ne reaction also contributes to the survival of $^{20}$Ne.}.
	
Observed and predicted classical nova abundances have been compared in many works over the years \citep[see, for example,][and references therein]{SS98,AK97,JJ98}. Models of classical novae for a range of ONe white dwarf masses have been presented in \citet{MP95}, but no attempt was made in that work to predict quantitatively the white dwarf mass of a given observed neon nova. It should also be mentioned that these models did not result in mass ejection, so their results must be treated with care. Nova model sequences have also been evolved by \citet{SW99}, with the aim of constraining ONe white dwarf masses of observed neon novae. However, this work does not present self-consistent models and must also be handled carefully. As will be seen later, some of their predictions differ vastly from our results. Clearly, the nova models presented in the latter two works are based on thermonuclear reaction rates that are about 20 years old and thus are completely outdated.

Furthermore, it must be emphasized that none of the studies to date explored the crucial effects of thermonuclear reaction rate variations on the predicted abundance ratios. A key aspect of our work is the use of a newly available nuclear physics library, called STARLIB \citep{CI11}. It incorporates a recent evaluation of thermonuclear reaction rates \citep{RL10,CI10a,CI10b,CI10c}, obtained from a Monte Carlo sampling of experimental nuclear physics observables, and provides for the first time statistically meaningful reaction rates and associated uncertainties.

In Sec. 2, we present updated observed abundances in neon nova shells. It will become clear that it is advantageous to focus the discussion on elemental abundance {\it ratios} rather than the abundance of one element. Our methods and results are presented in Secs. 3 and 4, respectively. First, we computed four new hydrodynamic models of classical novae, with peak temperatures in the range of 228-313 MK. The subset of elemental abundance ratios revealing the steepest dependence on peak temperature, that is, with the promise of potential use as a {\it thermometer}, was considered for further inspection. Then, suitable post-processing calculations were performed, demonstrating that the final predicted elemental abundances approximately agree with those from the self-consistent hydrodynamic model. In a final step, many post-processing network calculations were performed by varying one reaction rate at a time within its current uncertainty. The impact of this sensitivity study on potential classical nova thermometers was then investigated. Our results are compared to observation in Sec. 5. A summary and conclusions are given in Sec. 6.

\section{Observed Abundances\label{abund}} 
Before discussing our new simulations, we will summarize the abundances observed in neon nova shells. Compilations of classical nova abundances (expressed as mass fractions) can be found, for example, in \citet{RG98,SS98,SW99}. First, these compilations reveal that in several cases, even for identical events, different authors report rather different elemental abundances, with differences amounting up to an order of magnitude for some elements. Second, no uncertainties are provided with the listed abundances, making it difficult to assess the significance of a given element observation. Third, some of the compilations provide only the mass fraction sum for elements between Ne and Fe. As will become apparent later, when modeling neon novae the individual observed abundances of intermediate-mass elements (Ne, Mg, Al, Si,  and S) are extremely important. Fourth, for several novae listed in the above compilations, the mass fraction sum differs significantly from unity, reflecting errors in the reported values. Since all of these reasons could contribute to the mistaken impression that reported nova abundances are highly uncertain and scatter widely, we felt compelled to present updated abundances.

It is difficult to determine reliable abundances from observed nova ejecta. For example, the derived abundances are model-dependent since most analyses assume spherical shells although there is evidence that the ejecta are not spherically symmetric \citep{JC11}. Furthermore, the ejecta are potentially chemically inhomogeneous, adding substantially to the complexity of the problem. Also, the filling fraction (i.e., the fraction of the shell volume occupied by gas) is poorly constrained. Finally, the abundance analysis must account for the (sometimes substantial) fraction of unobserved ionization states. These, and other, difficulties have been discussed in more detail by \citet{JJ08}.

In the past different authors have used different procedures in the analysis of optical, infrared, and ultraviolet spectra. However, over the past 15 years many nova ejecta have been analyzed using consistent techniques, and the situation has improved significantly. In particular, we list all known neon novae in Table~\ref{abundances}, with the exception of V1370 Aql and U Sco\footnote{Abundance analyses of V1370 Aql are presented in \citet{MS87} and \citet{JA94}, but they were not performed using iterative least squares minimization techniques and hence the associated uncertainties are difficult to assess. For U Sco, only a few abundances, such as O and Ne, are discussed in \citet{EM11}.}. For the eight neon novae listed, the abundances were analyzed using the photoionization code CLOUDY \citep{GF03} coupled to the least squares minimization code MINUIT, adopted from the CERN library. In this manner, the intensities of all spectral lines are determined together via successive iterations during the minimization, while in addition uncertainties of the line intensities can be quantified in a meaningful way.

\begin{turnpage}
\centering
\begin{deluxetable*}{lcccccccccc}  % <--- column justification (center/left/right)
\tabletypesize{\scriptsize}
\tablecolumns{10}
\tablewidth{0pt}
%\rotate
\centering
\tablecaption{Summary of Observed Abundances (in Mass Fractions) for Neon Novae from UV, optical, and IR Spectroscopy\label{abundances}}
\tablehead{   % column headings
  \colhead{\phm{Ratios}} &
  \colhead{LMC 1990\#1\tablenotemark{1}} &
  \colhead{V4160 Sgr\tablenotemark{2}} &
  \colhead{V838 Her\tablenotemark{2}} &
  \colhead{V382 Vel\tablenotemark{3}} &
  \colhead{QU Vul\tablenotemark{4}} &
  \colhead{V693 CrA\tablenotemark{5}} &
  \colhead{V1974 Cyg\tablenotemark{6}} &
  \colhead{V1065 Cen\tablenotemark{7}} &
  \colhead{Solar\tablenotemark{8}}
}
\startdata
X$_{He}$/X$_H$		& 4.8(8)E-01			& 7.1(4)E-01			& 5.6(4)E-01		& 4.0(4)E-01			& 4.6(3)E-01			& 5.4(22)E-01		& 4.8(8)E-01		& 5.4(10)E-01		& 3.85E-01\\
X$_C$/X$_H$			& 3.7(15)E-02			& 1.43(7)E-02			& 2.28(23)E-02		& 2.6(13)E-03			& 9.5(59)E-04			& 1.06(44)E-02		& 3.1(9)E-03		& \nodata			& 3.31E-03\\
X$_N$/X$_H$			& 1.48(42)E-01			& 1.27(8)E-01			& 3.29(47)E-02		& 2.28(54)E-02			& 1.61(10)E-02			& 1.84(67)E-01		& 6.0(15)E-02		& 1.40(33)E-01 	& 1.14E-03\\
X$_O$/X$_H$			& 2.4(10)E-01			& 1.35(9)E-01		& 1.42(38)E-02		& 4.13(38)E-02			& 3.2(14)E-02			& 1.63(66)E-01		& 1.55(85)E-01		& 4.7(15)E-01		& 9.65E-03\\
X$_{Ne}$/X$_H$		& 1.6(10)E-01			& 1.38(5)E-01			& 1.22(5)E-01		& 4.0(7)E-02			& 5.1(4)E-02			& 6.7(34)E-01		& 9.7(40)E-02		& 5.34(98)E-01		& 2.54E-03\\
X$_{Mg}$/X$_H$		& 1.37(71)E-02			& $\approx$8.4E-03		& 1.2(7)E-03		& 2.45(14)E-03			& 1.02(49)E-02			& 9(7)E-03		& 4.3(28)E-03		& 4.4(13)E-02		& 9.55E-04\\
X$_{Al}$/X$_H$		& 2.3(11)E-02			& \nodata 				& 1.8(13)E-03		& 1.63(16)E-03			& 4.1(11)E-03			& 5.0(46)E-03		& $>$7.8E-05		& \nodata 			& 8.74E-05\\
X$_{Si}$/X$_H$		& 4.8(39)E-02			& 1.09(6)E-02			& 7(2)E-03		& 5(3)E-04			& 2.4(18)E-03			& 2.4(18)E-02		& \nodata			& \nodata			& 1.08E-03\\
X$_S$/X$_H$			& \nodata				& \nodata				& 1.48(15)E-02		&\nodata				&\nodata				& \nodata			& \nodata			& 2.3(13)E-02		& 5.17E-04\\
X$_{Ar}$/X$_H$		& \nodata				& \nodata				& \nodata			& \nodata				& 4.0(3)E-05			& \nodata			& \nodata			& 4.6(17)E-03 		& 1.29E-04\\
X$_{Fe}$/X$_H$		& \nodata 				& 2.4(8)E-03			& 2.35(63)E-03		& \nodata				& 9.53(54)E-04			& \nodata			& 8.8(72)E-03		& 1.16(40)E-02		 & 1.81E-03\\
X$_H$\tablenotemark{9}	& 4.7(9)E-01			& 4.65(37)E-01			& 5.63(36)E-01		& 6.6(4)E-01			& 6.3(3)E-01			& 3.8(14)E-01		& 5.5(8)E-01		& 3.6(10)E-01		& 7.11E-01\\
\enddata
\tablecomments{All abundances are given here in terms of mass fraction ratios, $X_{el}/X_H$ (or mass fraction for hydrogen; see last row), by converting the ``number abundances relative to hydrogen relative to solar" from the original literature (references provided below). The abundance uncertainties are given in parentheses.}
\tablenotetext{1}{From \citet{KV99}.}
\tablenotetext{2}{From \citet{GS07}.}
\tablenotetext{3}{From \citet{SS03}.}
\tablenotetext{4}{From \citet{GS02}.}
\tablenotetext{5}{From \citet{KV97}.}
\tablenotetext{6}{From \citet{KV05}; solar abundances assumed in their analysis are not listed; their adopted values were: $\log{(N_{el}/N_H)_\odot}$=$-1.0$ (He), $-3.45$ (C), $-4.03$ (N), $-3.13$ (O), $-3.93$ (Ne), $-4.42$ (Mg), $-5.53$ (Al), $-4.45$ (Si), $-4.79$ (S), $-4.49$ (Fe)\citep{KV12}.}
\tablenotetext{7}{From \citet{LH10}.}
\tablenotetext{8}{From \citet{KL09}.}
\tablenotetext{9}{Calculated from: $X_{H} = [1+ X_{He}/X_{H} + X{_C}/X{_H} + $\nodata$ + X_{Fe}/X_{H}]^{-1}$.}
\end{deluxetable*}
\end{turnpage}

Nova abundances are usually presented in the literature as ``relative to hydrogen relative to solar", i.e., the quantity
\begin{equation}
\xi=\frac{(N_{el}/{N_H})}{(N_{el}/{N_H})_\odot},
\end{equation}
where $N_{el}$ refers to the number abundance. However, the ``solar" abundances have undergone significant revision over the past two decades \citep[see discussion in][]{JJ08}, so observations were normalized to the solar abundances of \citet{KL09}. In order to compare the observations to our simulations (discussed below) in a meaningful way, we first calculated, from the values of $\xi$ reported in the literature, the quantity $N_{el}/N_H$ and, subsequently, converted the results to the elemental mass fraction ratio, $X_{el}/X_H$.  In some instances, the abundances presented in the literature had asymmetric uncertainties. For the sake of simplicity, the uncertainties in these cases were symmetrized by adopting the central value between the upper and lower bounds as the recommended value. The resulting mass fraction ratios for elements between He and Fe are listed in Table~\ref{abundances}. The hydrogen mass fraction is then obtained from
\begin{equation}
X_H=[1+\frac{X_{He}}{X_H}+\frac{X_C}{X_H}+...+\frac{X_{Fe}}{X_H}]^{-1}
\end{equation}
and is listed in the last row.

The individual mass fractions can be easily calculated from the values of $X_{el}/X_H$ and $X_H$. Note, however, that the values of $X_{el}$ derived in this manner are sensitive to the abundance fraction missed in the spectral line analysis, such as missing elements or unaccounted for ionization states. The ratio $X_{el}/X_H$ is expected to be less susceptible to systematic errors than $X_{el}$ and, therefore, we focus in the following sections on mass fraction {\it ratios} rather than mass fractions.

Inspection of Table~\ref{abundances} reveals large overabundances relative to the solar $X_{el}/X_H$ values, listed in the last column. For some neon novae, N, O, Ne, Mg, Al, Si, S\footnote{As will become apparent later, observations of sulfur are particularly important. In addition to the novae listed in Table~\ref{abundances}, sulfur has also been observed in the optical spectra of QV Vul, V2214 Oph and V443 Sct \citep{JA94}. For these three objects, which are most likely associated with outbursts involving CO rather than ONe white dwarfs, the derived sulfur abundances are near the solar value. Furthermore, \citet{JA94} report a relatively high sulfur abundance for QU Vul. However, the optical spectrum shown in their Figure 1 reveals only a weak sulfur [S II]  line. Their derived high abundance value is perhaps biased by their analysis method used. Finally, \citet{MS87} report a high sulfur abundance for the neon nova V1370 Aql. Unfortunately, no uncertainty is provided and the reliability of their reported value is difficult to assess.} and Ar, are enriched by factors of 160, 50, 260, 40, 260, 40, and 40, respectively. In the following sections, we will discuss the abundances predicted by models of neon novae.
 
\section{Method}\label{method}
Our nuclear reaction network followed the evolution of 117 nuclides in the range from H to Ti through 635 nuclear processes, including weak interactions, reactions of type (p,$\gamma$), (p,$\alpha$), ($\alpha$,$\gamma$), and their reverse reactions. The rates of these interactions are adopted from a next-generation library, called STARLIB \citep{CI11}. A detailed account of this library will be given in a forthcoming paper. In brief, STARLIB is a tabular library, listing the rates and corresponding uncertainties on a temperature grid between 1 MK and 10 GK for each nuclear interaction. For 62 of the nuclear reactions in the $A=14-40$ mass range contained in STARLIB (among them most of the reactions important for nova nucleosynthesis), {\it experimental} rates are based on the 2010 Monte Carlo evaluation of thermonuclear reaction rates \citep{RL10,CI10a, CI10b, CI10c}. The Monte Carlo reaction rates and associated uncertainties are obtained from sampling randomly over uncertainties of (input) nuclear physics observables and are derived from the 0.16th, 0.50th, and 0.84th quantiles of the cumulative distribution of the (output) reaction rates (for a coverage probability of 68\%). The information on rigorously defined  reaction rate uncertainties was not available previously and opens interesting possibilities for nucleosynthesis studies.

We generated a series of new hydrodynamic models using the code SHIVA \citep{JJ98} with reaction rates adopted from STARLIB. The different models were generated for a range of white dwarf masses (1.15 M$_{\odot}$ - 1.35 M$_{\odot}$). Information on derived parameters of our models, specifically, peak temperature ($T_{peak}$), ejected mass ($M_{ej}$), and initial white dwarf radius ($R_{WD}^{ini}$), is given in Table~\ref{nova}. Each model included 45 envelope zones. Test calculations performed with 500 zones provided essentially the same results. The peak temperatures achieved in the hottest zone in our simulations, ranging from 228 MK to 313 MK depending on the model, can be regarded as typical for models of thermonuclear runaways involving ONe white dwarfs \citep[for models that achieve higher peak temperatures, see][]{SS09}. We assumed a mixing fraction of 50\% between accreted matter of solar composition and white dwarf matter prior to the outburst. The solar abundances are adopted from \citet{KL09}, while the latter abundances are taken from the evolution of a 10 $M_{\odot}$ star from the main sequence to the end of core carbon burning \citep{CR96}. The initial envelope composition for our nova simulations is presented in Table~\ref{inicomp}. The initial luminosity and mass accretion rate for all models amounted to $L_{ini}=10^{-2}$ $L_\odot$ and $\dot{M}_{acc}=2\times10^{-10}$ M$_\odot /yr$, respectively.

\begin{deluxetable}{lcccc}
\tabletypesize{\small}
%\tablewidth{0pt}
\tablecaption{Properties of Evolutionary Nova Models\label{nova}}
\tablehead{
\colhead{Property} &  \multicolumn{4}{c}{Model}  \nl 
\cline{2-5}
&  \colhead{J115} & \colhead{J125} & \colhead{J130} & \colhead{J135}   
}
\startdata
M$_{WD}$ ($M_{\odot}$) & 1.15 & 1.25 & 1.30 & 1.35\\
$T_{peak}$ (GK) & 0.228 & 0.248 & 0.265 & 0.313\\
$M_{ej}$ ($10^{-5} M_{\odot}$) & 2.46 & 1.89 & 1.17 & 0.455\\
$R_{WD}^{ini}$ (km) & 4326 & 3788 & 3297 &2255\\
\enddata
\end{deluxetable}

\begin{deluxetable}{lc}
\tabletypesize{\small}
%\tablewidth{0pt}
\tablecaption{Initial Envelope Composition (Mass Fractions) of Present Nova Simulations\label{inicomp}}
\tablehead{
\colhead{Isotope} &  \colhead{Mass Fraction} }
\startdata
$^{1}$H 		& 3.56E-01\\
$^{3}$He  	& 4.23E-05\\
$^{4}$He  	& 1.37E-01\\
$^{6}$Li 		& 3.44E-10\\
$^{7}$Li  		& 4.91E-09\\
$^{9}$Be  		& 7.51E-11\\
$^{10}$B  		& 5.05E-10\\
$^{11}$B  		& 2.26E-09\\
$^{12}$C  	& 5.74E-03\\
$^{13}$C  	& 1.42E-05\\
$^{14}$N  	& 4.04E-04\\
$^{15}$N  	& 1.59E-06\\
$^{16}$O  	& 2.59E-01\\
$^{17}$O  	& 1.37E-06\\
$^{18}$O  	& 7.72E-06\\
$^{19}$F 		& 2.08E-07\\
$^{20}$Ne 	& 1.57E-01\\
$^{21}$Ne 	& 2.99E-03\\
$^{22}$Ne 	& 2.22E-03\\    
$^{23}$Na 	& 3.22E-02\\
$^{24}$Mg 	& 2.77E-02\\
$^{25}$Mg 	& 7.94E-03\\
$^{26}$Mg 	& 4.98E-03\\ 
$^{27}$Al 		& 5.43E-03\\
$^{28}$Si 		& 3.51E-04\\
$^{29}$Si 		& 1.85E-05\\
$^{30}$Si 		& 1.26E-05\\    
$^{31}$P  		& 3.50E-06\\
$^{32}$S  		& 1.74E-04\\
$^{33}$S  		& 1.42E-06\\
$^{34}$S  		& 8.24E-06\\     
$^{35}$Cl 	& 1.87E-06\\     
$^{37}$Cl 	& 6.29E-07\\
$^{36}$Ar 	& 3.84E-05\\
$^{38}$Ar 	& 7.40E-06\\
$^{39}$K  		& 1.86E-06\\
$^{40}$Ca 	& 3.18E-05\\
\enddata
\tablecomments{Values are obtained assuming 50\% mixing of solar accreted matter with white dwarf material (see text).}
\end{deluxetable}

Final isotopic abundances, for matter that reached and exceeded escape velocity (i.e., the fraction of the envelope effectively ejected), were determined 1 hour after peak temperature was achieved. By that time, short-lived parent nuclei had decayed to their stable daughters, and the isotopic abundances of a given element were summed for each model. Those {\it elemental} abundances revealing a strong, either monotonically increasing or decreasing, dependence on peak temperature were then considered for further study.

\begin{figure}
%%\epsscale{1}
\centering
\includegraphics[scale=0.45]{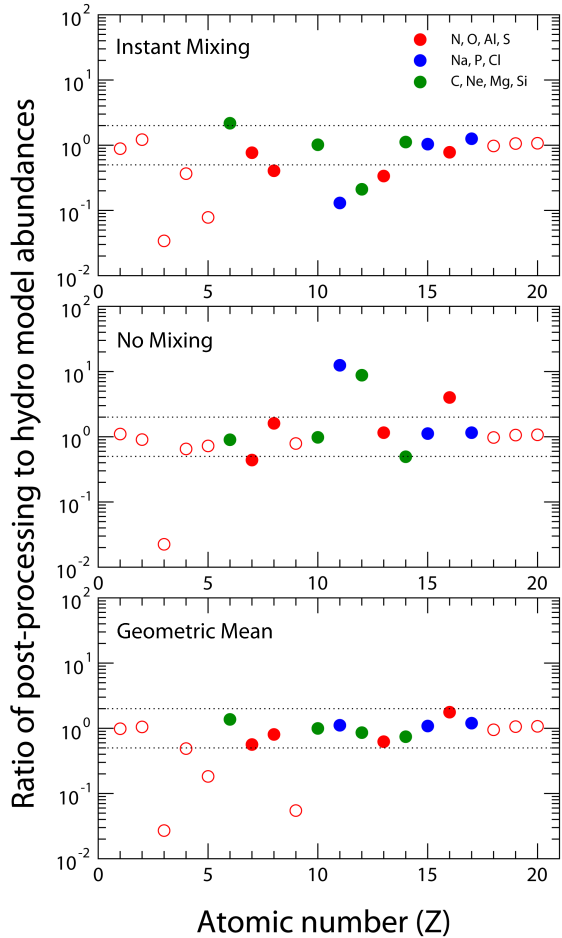}
%%\plotone{J125InstantAltpaper.eps}
\caption{(Color online) Comparison of final elemental abundances from a full hydrodynamic model to results from post-processing calculations. Abundance ratios are displayed versus atomic number, {\it Z}. The nova model assumes accretion onto a 1.25 M$_{\odot}$ ONe white dwarf (Table~\ref{nova}). The three panels are obtained for three different mixing assumptions in the post-processing calculations. {\bf Top Panel:} Instant-mixing; {\bf Middle Panel:} No-mixing; {\bf Bottom Panel:} Geometric mean of abundances from the top and middle panels. See text for details. Elements of interest in this work are displayed in color. Note the good agreement between hydrodynamic and post-processing results in the bottom panel. Similar plots are obtained for all other nova models explored in this work.\label{J125Instant}}
\end{figure}

Investigating the sensitivity of any useful nova thermometer to nuclear reaction rate variations implies that a large number of simulations need to be performed in order to exhaust the nuclear physics parameter space. Clearly, full hydrodynamic model simulations are at present far too time-consuming for this purpose. Instead, we performed post-processing reaction network calculations, using time-temperature-density profiles adopted from the hydrodynamic models described above. Such calculations have been used successfully in previous reaction rate sensitivity studies of classical novae \citep{CI02}. However, previous investigations employed one-zone calculations, appropriate for the hottest zone close to the white dwarf surface only. For the purposes of the present work it was important to perform post-processing calculations for {\it all burning zones}, matching the final abundances of the hydrodynamic models as closely as possible. We explored two extreme assumptions regarding the mixing of material between zones to this end. In the first case, we performed post-processing calculations for each burning zone separately, using the appropriate time-temperature-density profiles from the hydrodynamic study, and at the end of the calculation the elemental abundances were mass-averaged over all zones. This assumption is similar to the procedure used by \citet{MS02} and will be referred to as the ``no-mixing approximation". In the second case, the post-processing calculation is performed by replacing each local thermonuclear rate by its mass-weighted average over the convective region, implying that the turnover time is faster than the nuclear burning time. This assumption has been made, for example, by \citet{DP86} in the framework of a hydrodynamic model and will be referred to in the following as the ``instant-mixing approximation". The actual behavior probably lies somewhere between these prescriptions. Therefore, a third set of elemental abundances was derived from the post-processing calculations, simply by computing the geometric mean of the abundances resulting from the no-mixing and instant-mixing approximations. Representative results are shown in Figure~\ref{J125Instant}, comparing the final elemental abundances from a full hydrodynamic simulation with the three sets of final post-processing abundances described above. It is especially encouraging that the geometric mean abundance values seem to reproduce the stellar model elemental predictions rather well especially for the elements of interest in this work.

More than 7,000 multi-zone network calculations for all four classical nova hydrodynamic models were performed by independently varying the rates of 214 reactions, that is, the most important nuclear reactions in our network. For each reaction, the rates were multiplied by factors of 100, 10, 5, 2, 0.5, 0.2, 0.1, and 0.01 in successive network calculations. Final elemental abundances were adopted from the mixing prescription (instant-mixing, no-mixing, or geometric mean) that best approximated the results of a given hydrodynamic model using recommended reaction rates. Final elemental abundances are then linearly interpolated between the above variation factors to determine the impact of the actual reaction rate uncertainties, which were adopted from the STARLIB library.

\section{Results}\label{results}
\begin{figure}[t]
%%\epsscale{1.1}
\centering
\includegraphics[scale=0.45]{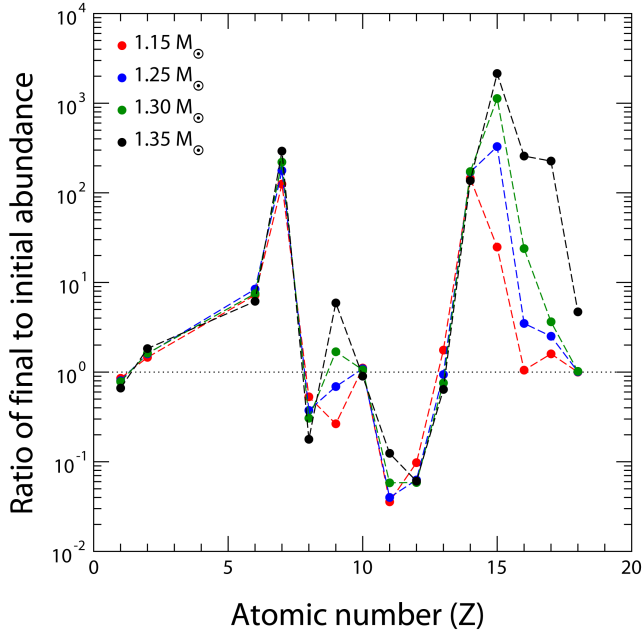}
%%\plotone{hydroElementAbundAltpaper.pdf}
\caption{(Color online) Final elemental abundances from all four hydrodynamic models explored in this work, normalized to initial abundances, versus atomic number. Some elements are overproduced (N, Si, P, S), while others are underproduced (O, Na, Mg). Of interest here is the monotonic dependence of certain elements on the white dwarf mass or, equivalently, peak temperature, resulting in a color sequence of red, blue, green, black, or vice versa. \label{hydroElementAbund}}
\end{figure}

We start with the results from our hydrodynamic simulations using the code SHIVA. Figure~\ref{hydroElementAbund} displays the final elemental abundances, normalized to initial abundance adopted in the simulation, versus atomic number, $Z$, for all four nova models (see Table~\ref{nova}). It is apparent that certain elements are overproduced (N, Si, P, and S), while others are depleted (O, Na, Mg). We are particularly interested in final elemental abundances that show a steep dependence on peak temperature. We find that N, Na, P, and S monotonically increase with peak temperature while O and Al decrease. Note that F, Cl, and Ar also show strong trends, but their absolute predicted final abundances are very small (i.e., 10$^{-7}$ to 10$^{-5}$ by mass). Therefore, they were deemed not useful for our present purpose. Based on the hydrodynamic simulations, our first indication of useful thermometers, with a steep dependence on peak temperature, are the eight element ratios N/O, N/Al, O/S, S/Al, O/Na, Na/Al, O/P, and P/Al.

\begin{figure}[h]
%%\epsscale{1.1}
\centering
\includegraphics[scale=0.45]{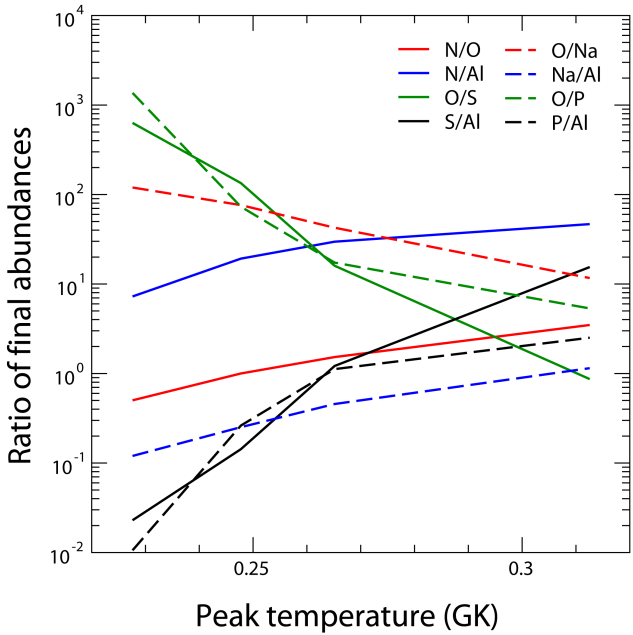}
%%\plotone{ratiosAlt2.eps}
\caption{(Color online) Ratios of 8 elemental abundances (mass fractions), derived from hydrodynamic nova models, that show a steep and monotonic dependence on peak temperature, i.e., by at least an order of magnitude. These ratios are prime candidates for {\it nova thermometers}.\label{ratios}}
\end{figure}

The abundance ratios for these element pairs are shown in Figure~\ref{ratios} versus peak temperature, where the solid lines indicate ratios for elements that have been observed in neon nova shells (N, O, Al, S), and dashed lines denote ratios involving elements (Na, P) for which reliable abundances have not been reported in shells of classical novae yet\footnote{Infrared coronal lines from phosphorus, [P VIII] and [P VII], in the ejecta of V1974 Cyg have been reported by \citet{RW96}, but no elemental abundances have been reported.}. The elemental abundance ratios O/S (decreasing) and S/Al (increasing) display the strongest variation between peak temperatures of 228 MK and 313 MK, amounting to about three orders of magnitude. Following closely are the abundance ratios O/P (decreasing) and P/Al (increasing), which vary by more than two orders of magnitude. The other ratios shown in Figure~\ref{ratios}  (O/Na, N/Al, N/O, Na/Al) show variations by about one order of magnitude. The next step was to investigate how robust these element ratios are with regard to thermonuclear reaction rate variations.

\begin{figure*}
%%\epsscale{1.8}
\centering
\includegraphics[scale=0.55]{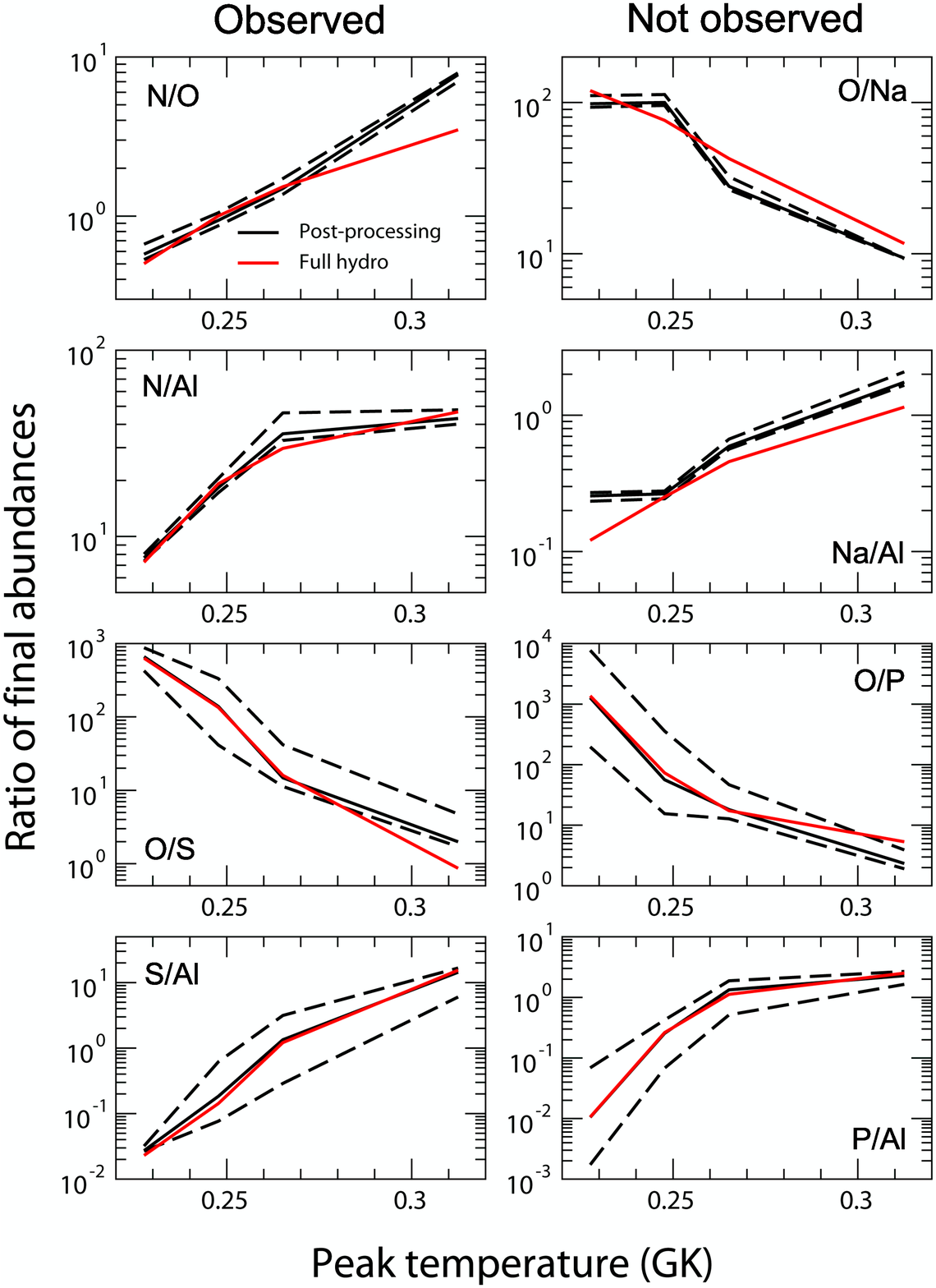}
%%\plotone{eightpanel5.eps}
\caption{(Color online) Eight Ratios of elemental abundances (mass fractions), prime candidates for {\it nova thermometers}, that show a steep and monotonic dependence on peak temperature. Solid black and red lines correspond to element ratios obtained from our post-processing and hydrodynamic simulations, respectively. Dashed lines indicate the uncertainty bands obtained by independently varying all relevant nuclear reaction rates within their uncertainties. Broad uncertainty bands call for future laboratory measurements of nuclear reactions (Table~\ref{reactions}).
{\bf Left Panel:} All ratios shown in the left column involve elements that have been {\it observed} in neon nova shells (see Table~\ref{abundances}).
{\bf Right Panel:} All ratios shown in the right column involve elements (Na and P) that have {\it not been observed} in neon nova shells (see Table~\ref{abundances}).\label{eightratios1}}
\end{figure*}

The sensitivity of the nova thermometers to the nuclear physics input is shown in Figure~\ref{eightratios1}. Elemental abundance (mass fraction) ratios are displayed versus peak temperature, for the four nova models explored here (see Table~\ref{nova}). Shown on the left-hand side are abundance ratios involving elements that have been observed in neon nova shells, while the right-hand side displays abundance ratios involving at least one element that has not yet been observed. The solid red lines correspond to final abundance ratios obtained from the full hydrodynamic models, while the solid black lines show the results of the mixing prescription that best reproduced the hydrodynamic model results using recommended reaction rates. Note that the red and black solid lines generally differ in magnitude by less than a factor of 2, providing support for our conjecture that the present post-processing simulations reliably approximate the results from the full hydrodynamic models for the elements of interest in this work.

The dashed lines in each panel of Figure~\ref{eightratios1} display the change of predicted elemental abundance (mass fraction) ratios caused by varying individual reaction rates within their uncertainties \citep[defined by a coverage probability of 68\%, see][]{RL10,CI10a}. Interestingly, four of the thermometers (N/O, N/Al, O/Na, and Na/Al) reveal rather robust abundance ratios, with an uncertainty of less than 30\%. Consequently, these ratios represent currently useful nova thermometers, without the immediate need for improved nuclear laboratory measurements. However, only the first two (N/O and N/Al) involve elements that have so far been observed in neon nova shells. The last two involve sodium, which has been searched for in infrared spectra but not yet detected \citep{SS12}.

The use of the other four predicted elemental ratios (O/S, S/Al, O/P, and P/Al) as nova thermometers is somewhat limited. The first two of these ratios (O/S and S/Al) involve elements observed in neon nova shells (Table~\ref{abundances}), while the other two ratios (O/P and P/Al) involve phosphorous, which has not been observed. Thus, our results call for improved observations and spectral analysis of neon nova ejecta (i.e., for phosphorous). These ratios are also uncertain by factors of 3-6, so future laboratory measurements of nuclear reactions would clearly improve their utility as nova thermometers. However, these four ratios also demonstrate the strongest temperature dependence of the potential thermometers, which mitigates the effect of reaction rate uncertainties. Therefore, the two currently observed ratios (O/S and S/Al), despite their large uncertainties, can be effectively used to narrow the range of white dwarf mass for a nova, as will be shown in Section~\ref{obs} for V838 Her (Figure~\ref{v838her}).

Detailed information on our results is provided in Table~\ref{reactions}. The first four entries (N/O, N/Al, O/S, and S/Al) relate to nova thermometers involving elements observed in neon nova shells, while the last four involve sodium and phosphorous which have not been observed. Column 2 lists the maximum range of a given elemental abundance ratio versus peak temperature from our post-processing analysis. Recall that the larger this value (i.e., the steeper the dependence on temperature), the more useful a ratio will be as a nova thermometer.

The next columns list the two reactions whose current rate uncertainties most significantly influence a given elemental abundance ratio. A number of important conclusions can be drawn from this information. First, the ``second-most important reactions" (columns 5 and 6) give rise to very small elemental abundance ratio variations (less than 30\%), a number that is small compared to the uncertainty in observed elemental abundances (Table~\ref{abundances}). Second, the nova thermometers O/S, S/Al, O/P, and P/Al, which have the steepest dependence on peak temperature (by factors of 220-540; see column 2) also show the largest sensitivity to current reaction rate uncertainties (by factors of 3-6; see column 4). Third, the latter four nova thermometers are mainly sensitive to current rate uncertainties of a single reaction, $^{30}$P(p,$\gamma$)$^{31}$S. This reaction involves a short-lived nuclide (t$_{1/2}$=2.498 min) and has not been measured directly yet because a sufficiently intense $^{30}$P beam is currently lacking. Indirect nuclear structure studies, in order to improve the reaction rate, have been reported \citep{DD12, AP11, CW09}. At present, the spin and parity assignments for some of the important threshold states in the $^{31}$S compound nucleus are ambiguous. In addition, none of the proton partial widths for these levels are known experimentally. Therefore, we did not estimate the rates using the Monte Carlo procedure described in Sec. 3. In the absence of more reliable information, we adopted the Hauser-Feshbach (statistical model) estimate of \citet{TR00}, assuming a factor of 10 uncertainty in the classical nova temperature range. Our adopted rates for this reaction agree within their uncertainties with those of \citet{AP11}, which were obtained using a different procedure.

\begin{deluxetable*}{lccccccc}
\tabletypesize{\small}
\tablewidth{0pt}
\tablecaption{Uncertainty Sources of Predicted Elemental Abundance Ratios in Neon Nova Shells\label{reactions}}
\tablehead{
\colhead{Ratio} & \colhead{Range\tablenotemark{1}}  & \multicolumn{2}{c}{Primary Source} & \colhead{\phm{1}} & \multicolumn{2}{c}{Secondary Source}\nl 
\cline{3-4}
\cline{6-7}
&  \phantom & \colhead{Reaction} & \colhead{Uncertainty\tablenotemark{2}} & \colhead{\phm{1}} & \colhead{Reaction} & \colhead{Uncertainty\tablenotemark{2}}  
}
\startdata
N/O & 13.4 & $^{16}$O(p,$\gamma$)$^{17}$F & 1.16 & \phantom & $^{13}$N(p,$\gamma$)$^{14}$O & 1.06\\
N/Al & 5.59 & $^{20}$Ne(p,$\gamma$)$^{21}$Na & 1.29 & \phantom & $^{13}$N(p,$\gamma$)$^{14}$O & 1.18\\
O/S & 332 & $^{30}$P(p,$\gamma$)$^{31}$S & 3.36 & \phantom & $^{28}$Si(p,$\gamma$)$^{29}$P & 1.09\\
S/Al & 529 & $^{30}$P(p,$\gamma$)$^{31}$S & 4.62 & \phantom & $^{28}$Si(p,$\gamma$)$^{29}$P & 1.12\\
\tableline
O/Na & 10.8 & $^{16}$O(p,$\gamma$)$^{17}$F & 1.16 & \phantom & $^{20}$Ne(p,$\gamma$)$^{21}$Na & 1.12\\
Na/Al & 6.83 & $^{23}$Na(p,$\gamma$)$^{24}$Mg & 1.19 & \phantom & $^{20}$Ne(p,$\gamma$)$^{21}$Na & 1.10\\
O/P & 541 & $^{30}$P(p,$\gamma$)$^{31}$S & 6.44 & \phantom & $^{16}$O(p,$\gamma$)$^{17}$F & 1.26\\
P/Al & 216 & $^{30}$P(p,$\gamma$)$^{31}$S & 6.53 & \phantom & $^{20}$Ne(p,$\gamma$)$^{21}$Na & 1.22\\
\enddata
\tablenotetext{1}{Factor variation of final elemental abundance ratio over range of nova models explored in present work, obtained from the solid black lines shown in Figure~\ref{eightratios1}.}
\tablenotetext{2}{Factor uncertainty of final elemental abundance (mass fraction) ratio, caused by varying rate of individual reaction within its current uncertainty.}
\end{deluxetable*}

The influence of the $^{30}$P(p,$\gamma$)$^{31}$S rate on the abundances in the mass region above Si has already been pointed out by \citet{JJ01} and, more recently, by \citet{AP11}. We have quantified its impact on nova thermometers in the present work. Therefore, our results provide additional motivation for new laboratory measurements at radioactive ion beam facilities.

\begin{figure}[h]
%%\epsscale{1.3}
\centering
\includegraphics[scale=0.4]{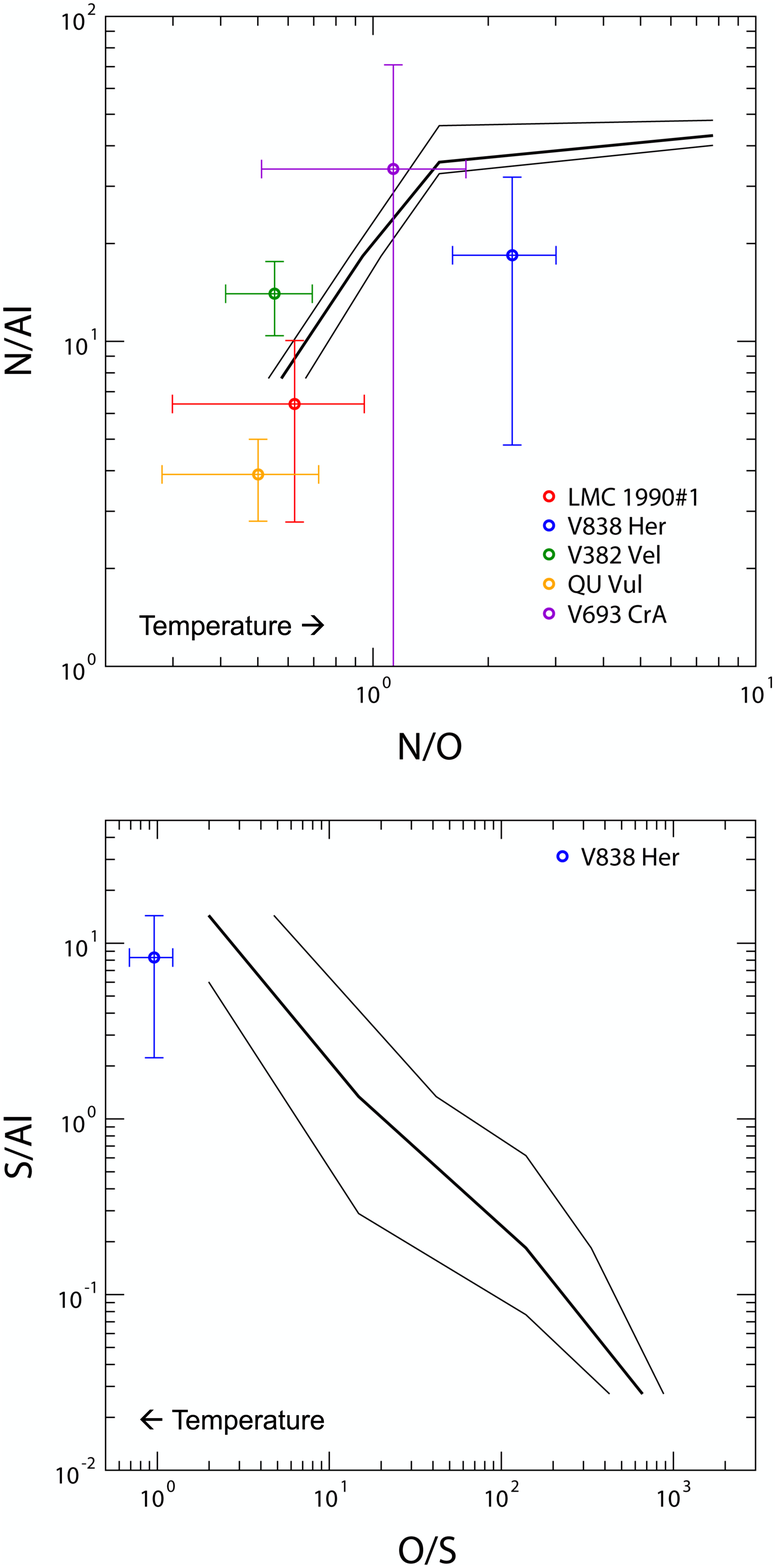}
%%\plotone{combined.eps}
\caption{(Color online) Three-element plots used to compare elemental ratios predicted by post-processing calculations to that of classical nova observations. The thick, solid black line corresponds to the elemental ratios predicted by post-processing calculations using recommended reaction rates. The thinner black lines denote the range of uncertainty in these ratios due to reaction rate variations (see Figure~\ref{eightratios1}). The data points give the elemental ratios as observed in several neon novae (see Table~\ref{abundances}). {\bf Top Panel:} Three-element plot of N, O, and Al.  The peak temperature in the simulation results increases from left to right. {\bf Bottom Panel:} Three-element plot of O, Al, and S. Note that peak temperature in the simulation results increases from right to left in this figure.\label{NAl-NO}}
\end{figure}

In general, two different techniques can be used to study the impact of reaction-rate variations on nucleosynthesis: variation of individual rates, as done in the present work, and simultaneous variation of all rates using a Monte Carlo approach. The advantage of the first method is that the impact of a specific reaction can be quantified in a straightforward manner, whereas the second technique allows for the influence of correlations between interactions in the network to be taken into account automatically. While our results indicate that primary sources of uncertainty, such as uncertainty in the $^{30}$P(p,$\gamma$)$^{31}$S rate, overwhelm the effects of any secondary source of uncertainty, we chose to verify our individual variation method. A recently implemented Monte Carlo reaction network, which included the instant-mixing and no-mixing approximations, was used to reproduce the nuclear thermometers and their uncertainties. This approach supports the results of individual reaction rate variation when correlation effects are taken into account, similar to the work of \citet{AP08}.

\section{Comparison with Observation}\label{obs}
We will now compare the abundances from our simulations with observations (Table~\ref{abundances}). A first general impression can be obtained from the top panel of Figure~\ref{NAl-NO}, showing a three-element abundance plot involving N, O and Al, i.e., N/O vs. N/Al. Our model predictions are shown as thick and thin solid, black lines (see Figure~\ref{eightratios1}). As already discussed in Sec. 4, the simulated N/O and N/Al abundance ratios have relatively small uncertainties,  which is reflected by the narrow region between the two thin solid lines. These abundance ratios represent useful thermometers since they are not significantly affected by current reaction rate uncertainties. The elements N, O, and Al have been simultaneously observed in five neon novae, according to Table~\ref{abundances}, and their abundance ratios are displayed as data points. It is apparent that in this case the uncertainties in the observational data, with the exception of V382 Vel (green), far exceed those in the simulations. We conclude that the N/O and N/Al abundance ratios will become more useful thermometers for a number of novae when more reliable abundances can be determined from UV, IR, and optical spectra. A three-element abundance plot involving O, Al, and S, i.e., O/S vs. S/Al, is shown in bottom panel of Figure~\ref{NAl-NO}. We previously mentioned in Sec. 4 that the predicted O/S and S/Al abundance ratios have rather large uncertainties, which is reflected by the broad region between the two thin, solid lines. These uncertainties originate from the poorly known $^{30}$P(p,$\gamma$)$^{31}$S reaction rate (Figure~\ref{eightratios1} and Table~\ref{reactions}). Only for a single neon nova, V838 Her (blue), have all three elements been observed simultaneously (Table~\ref{abundances}). It is apparent in this case that the observational uncertainties are relatively small compared to the stellar model results. We conclude that the O/S and S/Al abundance ratios will become more useful thermometers for neon novae when the $^{30}$P(p,$\gamma$)$^{31}$S reaction rate can be determined experimentally and Al and S are simultaneously observed in additional novae.

A more detailed comparison between prediction and observation can be made by comparing observed abundance ratios to two or more corresponding nuclear thermometers. We are interested to find out if, for a given neon nova, the derived peak temperatures from two or more nuclear thermometers are in mutual agreement. Only Nova V838 Her exhibits observed elemental abundance ratios corresponding to all four currently observable nuclear thermometers (N/O, N/Al, O/S, and S/Al), while observed abundance ratios corresponding to two nuclear thermometers (N/O and N/Al or N/O and O/S) are available for five other neon novae (Table~\ref{abundances}). Thus we are disregarding in the following discussion neon novae V4160 Sgr and V1974 Cyg, for which only a single nuclear thermometer (N/O) is available.

\begin{figure}
%%\epsscale{1.7}
\centering
\includegraphics[scale=0.65]{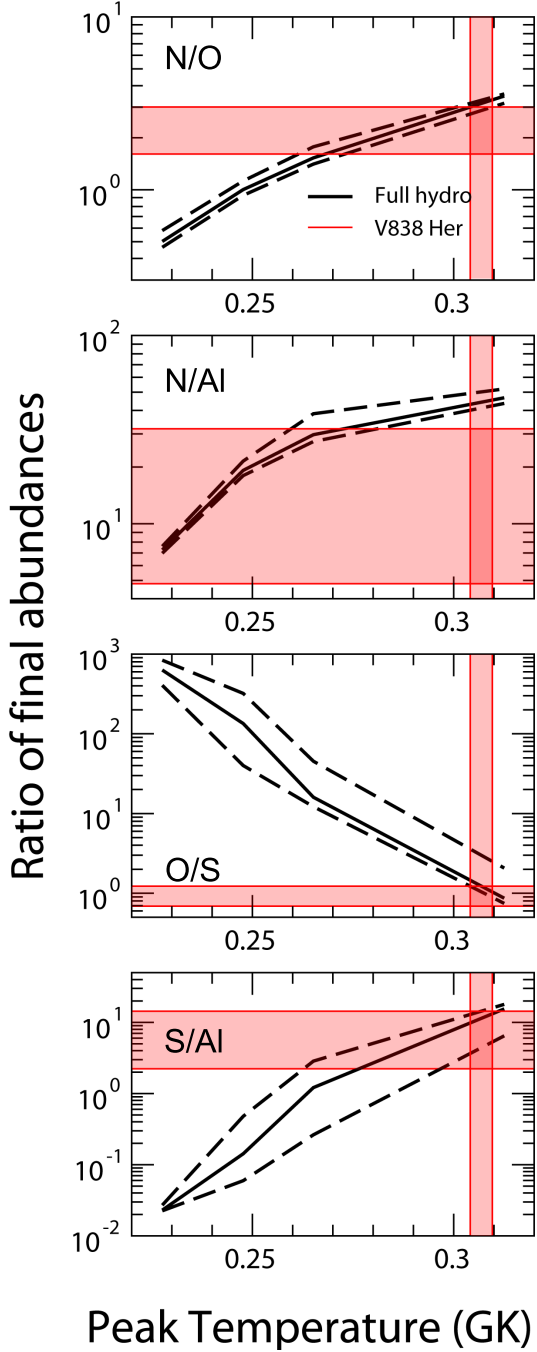}
%%\plotone{fourpanelwithwdmass.eps}
\caption{(Color online) Detailed comparison between prediction and observation of Nova V838 Her for all four currently observable nova thermometers (N/O, N/Al, O/S, and S/Al). The solid black lines correspond to elemental ratios obtained from hydrodynamic simulations. The dashed lines indicate the uncertainty bands which were obtained by post-processing and extrapolated to the hydrodynamic results. The horizontal red bands correspond to the range of observed values of V838 Her for each elemental ratio. The vertical red bands show the estimated peak temperature range for this nova based on the intersection of predicted and observed values.
\label{v838her}}
\end{figure}

{\it Nova V838 Her:} Predicted and observed values for Nova V838 Her are presented in Figure~\ref{v838her}. The observed N/O and O/S abundance ratios strongly restrict the peak temperature to a range of $T_{peak}=0.30-0.31$ GK, corresponding to a white dwarf mass of $M_{WD}=1.34-1.35$ M$_{\odot}$. For this temperature range, the observed N/Al abundance ratio barely misses the predicted values, although it is clear that the uncertainty of the observed value is too large for the observation to be of any use. An improved estimate of the white dwarf mass range could be obtained by reducing the nuclear uncertainties in the O/S and S/Al thermometers.

Our above estimate supports the suggestion of \citet{MP95} that the white dwarf in this system ``is very massive." On the other hand, our estimate significantly exceeds the result of \citet{SW99}, which suggests a value of $M_{WD}=1.05$ M$_{\odot}$. Notice that our simulations reproduce the observed abundance ratios in the narrow peak temperature range shown in Figure~\ref{v838her}, although no ``breakout" of matter from the CNO mass range to heavier masses has occurred in any of our nova models. Thus we find no compelling evidence for a breakout in V838 Her, contrary to previous results \citep{GS07}.

{\it Nova V382 Vel:} The N/O and N/Al  thermometers, which exhibit relatively small nuclear uncertainties, restrict the peak temperature to a narrow range of $T_{peak}=0.23-0.24$ GK, corresponding to a white dwarf mass of $M_{WD}=1.18-1.21$ M$_{\odot}$. For this relatively low peak temperature range, only a modest overproduction (factor 2) of sulfur is expected (Figure~\ref{hydroElementAbund}), consistent with its lack of observation in this neon nova.

{\it Nova V693 CrA:} The observed N/O and N/Al abundance ratios have relatively large uncertainties and, consequently, the peak temperature is not well constrained. The best estimate, based on these two nuclear thermometers, results in an upper limit of $T_{peak} \leq 0.28$ GK, corresponding to a white dwarf mass of $M_{WD} \leq 1.3$ M$_{\odot}$. An improved estimate could be obtained with more reliable observed abundance ratios. Our result is significantly larger than that of \citet{SW99}, who suggest a value of $M_{WD}=1.05$ M$_{\odot}$.

{\it Nova LMC 1990\#1:} The uncertainties of the observed N/O and N/Al elemental abundance ratios are too large to narrowly constrain the peak temperature range, and thus we only obtain an upper limit of $T_{peak} \leq 0.24$ GK, corresponding to a white dwarf mass of $M_{WD} \leq 1.2$ M$_{\odot}$. Improved observations would certainly result in a more reliable prediction.

{\it Nova QU Vul:} The observed N/O ratio hints at a low peak temperature. However, the observed N/Al abundance ratio is smaller than the predicted values over the entire peak temperature range explored in the present work. Therefore, we can only conclude that the peak temperature is $T_{peak} < 0.22$ GK, corresponding to a white dwarf mass of $M_{WD} < 1.2$ M$_{\odot}$. This value is near the {\it lower limit} for single ONe white dwarfs \citep{CD10}. It also agrees with the suggestion of \citet{SW99}, who report a range of $M_{WD}=1.05-1.1$ M$_{\odot}$.

{\it Nova V1065 Cen:} In this case, observed N/O and O/S abundance ratios are available. However, no range of peak temperatures can be found that give consistent results for these two nuclear thermometers.

\begin{figure}[t!]
\centering
\includegraphics[scale=0.4]{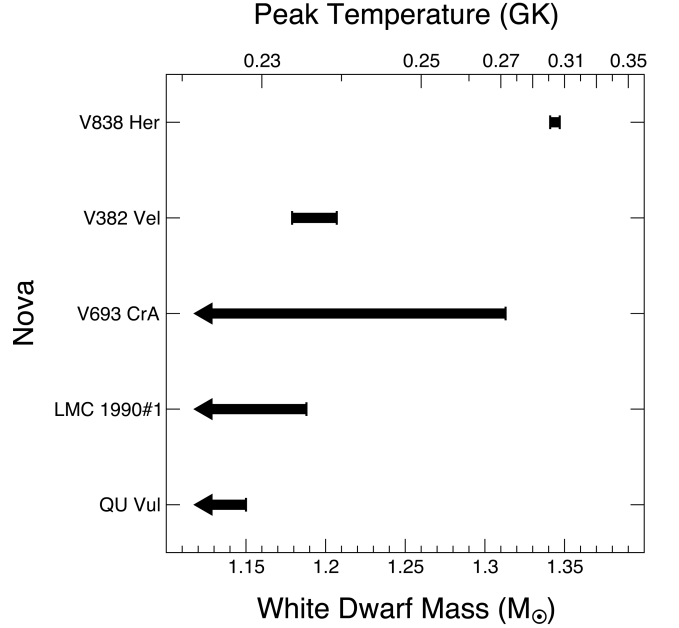}
%%\epsscale{1.05}
%%\plotone{bargraphpaper.eps}
\caption{(Color online) Peak temperature and white dwarf mass ranges as given by intersection of observed and predicted values of nova thermometers (see, for example, Figure~\ref{v838her}). Bracketed bars refer to white dwarf mass ranges where a definite range could be established. Bars terminating with an arrow refer to white dwarf mass ranges where only an upper limit could be determined.
\label{wddetermination}}
\end{figure}

The resulting peak temperature and white dwarf mass ranges for V838 Her, V382 Vel, V693 CrA, LMC 1990\#1, and QU Vul are shown in Figure~\ref{wddetermination}. Although these results are encouraging, it must be kept in mind that the present simulations are obtained for one particular choice of accretion rate, white dwarf luminosity, and mixing fraction between accreted and white dwarf matter. Evolutionary sequences for other choices of these parameters are in progress and will be presented in a forthcoming publication.

\section{Summary and Conclusions}\label{summary}
One goal of the present work was to investigate how useful elemental abundances are for determining the peak temperature, which is strongly correlated with the underlying white dwarf mass, achieved during neon nova outbursts. Another aim was to determine how robust such ``nova thermometers" are with respect to currently uncertain nuclear physics input. To this end, we first presented updated observed abundances for several neon novae, and we performed new hydrodynamic simulations of neon novae, with peak temperatures in the range of 228 MK to 313 MK. We found that the most useful thermometers (i.e., those elemental abundance ratios with the steepest monotonic dependence on peak temperature) are N/O, N/Al, O/S, S/Al, O/Na, Na/Al, O/P, and P/Al. These vary by factors of 13, 6, 330, 530, 11, 7, 540, and 220, respectively, over the temperature range explored here (see Table~\ref{reactions}). 

Next, we investigated the sensitivity of these nova thermometers to thermonuclear reaction rate variations. The ratios N/O, N/Al, O/Na, and Na/Al are robust, in the sense that they are affected by uncertain reaction rates by less than 30\%. At present, only the ratios N/O and N/Al are useful thermometers since these elements have been observed in neon novae (see Table~\ref{abundances}). The last two ratios, O/Na and Na/Al, are not currently useful as sodium has not yet been detected in classical nova spectra. We also find that for these four elemental abundance ratios the dependence on peak temperature (factor variations of 6-13; Table~\ref{reactions}) is not as strong as for some other ratios, mentioned below.

The thermometers O/S, S/Al, O/P, and P/Al reveal far steeper monotonic dependences on peak temperature, by factors of 220-540. However, their current drawback is the strong dependence on the uncertain $^{30}$P(p,$\gamma$)$^{31}$S reaction rate. Thus, our study provides additional motivation for new laboratory measurements of this crucial nuclear reaction. In addition, the last two of these ratios, O/P and P/Al, involve phosphorus, another element for which reliable abundances have not yet been derived from classical nova spectra in outburst.

Finally, we compared our model predictions to elemental abundances observed in neon nova shells. Based on the present nuclear thermometers, we obtain the following estimates for the underlying ONe white dwarf masses: 1.34-1.35 M$_{\odot}$ (V838 Her), 1.18-1.21 M$_{\odot}$ (V382 Vel), $\leq$1.3 M$_{\odot}$ (V693 CrA), $\leq$1.2 M$_{\odot}$ (LMC 1990\#1), and $\leq$1.2 M$_{\odot}$ (QU Vul). Presently no white dwarf mass range based on elemental abundances can be derived for nova V1065 Cen. These predictions could be improved if more reliable abundances from the analysis of UV, IR, and optical spectra become available in the future. The present simulations were obtained for one particular choice of accretion rate, white dwarf luminosity, and mixing fraction between accreted and white dwarf matter. Evolutionary sequences for other choices of these parameters will be presented in a forthcoming publication.

We would like to thank Anne Sallaska, Greg Schwarz, Steve Shore, and Karen Vanlandingham for fruitful discussions. This work was supported in part by the US Department of Energy under grant DE-FG02-97ER41041, the National Science Foundation under award number AST-1008355, the Spanish MICINN grants AYA2010-15685 and EUI2009-04167, the E.U. FEDER funds, the ESF EUROCORES Program EuroGENESIS, and by Arizona State University funding from the National Science Foundation and the National Aeronautics and Space Administration.

\clearpage
\end{document}